\documentclass[11pt]{article}
\usepackage{color,fullpage,url,booktabs}
\usepackage[colorlinks,pdftex]{hyperref} 
\usepackage{amsmath, amsthm, amssymb}  
\newcommand{\nc}{\newcommand}
\nc{\rnc}{\renewcommand}

\DeclareMathOperator{\diag}{diag}

\DeclareMathOperator{\poly}{poly}

\DeclareMathOperator{\tr}{tr}

\DeclareMathOperator{\Var}{Var}

\def\be#1\ee{\begin{equation}#1\end{equation}}
\def\bea#1\eea{\begin{eqnarray}#1\end{eqnarray}}
\def\beas#1\eeas{\begin{eqnarray*}#1\end{eqnarray*}}
\def\ba#1\ea{\begin{align}#1\end{align}}
\def\bas#1\eas{\begin{align*}#1\end{align*}}
\def\bpm#1\epm{\begin{pmatrix}#1\end{pmatrix}}

\def\eq#1{\eqref{eq:#1}}

\def\ra{\rightarrow}

\nc{\grad}{{\vec{\nabla}}}

\newtheorem{thm}{Theorem}
\newtheorem*{thm*}{Theorem}

\newtheorem{proto}{Protocol}

\theoremstyle{definition}

\newtheorem{dfn}[thm]{Definition}
\theoremstyle{plain}

\makeatletter
\newtheorem*{rep@theorem}{\rep@title}
\newcommand{\newreptheorem}[2]{%
\newenvironment{rep#1}[1]{%
 \def\rep@title{#2 \ref{##1} (restatement)}%
 \begin{rep@theorem}}%
 {\end{rep@theorem}}}
\makeatother

\newreptheorem{thm}{Theorem}
\newreptheorem{lem}{Lemma}

\def\cE{\mathcal{E}}

\def\cO{{\cal O}}
\def\cP{\mathcal{P}}

\def\cZ{\mathcal{Z}}

\def\bbR{\mathbb{R}}

\def\benum{\begin{enumerate}}
\def\eenum{\end{enumerate}}
\def\bit{\begin{itemize}}
\def\eit{\end{itemize}}

\newcommand{\secref}[1]{Section~\ref{sec:#1}}

\newcommand{\lemref}[1]{Lemma~\ref{lem:#1}}
\newcommand{\thmref}[1]{Theorem~\ref{thm:#1}}

\nc{\todo}[1]{\textcolor{red}{todo: #1}}

\def\begsub#1#2\endsub{\begin{subequations}\label{eq:#1}\begin{align}#2\end{align}\end{subequations}}
\nc\qand{\qquad\text{and}\qquad}
\nc\mnb[1]{\medskip\noindent{\bf #1}}

\nc{\pder}[2]{\frac{\partial {#1}}{\partial {#2}}}
\nc{\p}{\partial}

\usepackage{listings}
\usepackage{graphicx}
\usepackage{tikz}
\usepackage{float}
\newtheorem{theo}[thm]{Theorem}

\newtheorem{lemma}[thm]{Lemma}

\usepackage{boxedminipage}

\usepackage{bbm}
\usepackage{yfonts}
\usepackage{extarrows}
\usepackage{mathrsfs}
\newcommand{\ST}{\textrm{ST}}
\newcommand{\mix}{\textrm{mix}}
\newtheorem*{dfn*}{Definition}
\begin{document}
\title{Simulated Quantum Annealing Can Be Exponentially Faster than Classical Simulated Annealing}
\author{Elizabeth Crosson\thanks{Caltech IQIM. {\tt crosson@caltech.edu}} \and Aram W. Harrow\thanks{MIT CTP. {\tt aram@mit.edu}}}
\maketitle

\abstract{
Can quantum computers solve optimization problems much more quickly than classical computers?  One major piece of evidence for this proposition has been the fact that Quantum Annealing (QA, also known as adiabatic optimization) finds the minimum of some cost functions exponentially more quickly than Simulated Annealing (SA), which is arguably a classical analogue of QA.  

One such cost function is the simple ``Hamming weight with a spike'' function in which the input is an $n$-bit string and the objective function is simply the Hamming weight, plus a tall thin barrier centered around Hamming weight $n/4$.   While this problem can be solved by inspection, it is also a plausible toy model of the sort of local minima that arise in real-world optimization problems.  It was shown by Farhi, Goldstone and Gutmann~\cite{farhi-2002} that for this example SA takes exponential time and QA takes polynomial time, and the same result was generalized by Reichardt~\cite{Reichardt-2004} to include barriers with width $n^\zeta$ and height $n^\alpha$ for $\zeta + \alpha \leq 1/2$.  This advantage could be explained in terms of quantum-mechanical ``tunneling.''

Our work considers a classical algorithm known as Simulated Quantum Annealing (SQA) which relates certain quantum systems to classical Markov chains.  By proving that these chains mix rapidly, we show that SQA runs in polynomial time on the Hamming weight with spike problem in much of the parameter regime where QA achieves exponential advantage over SA.   While our analysis only covers this toy model, it can be seen as evidence against the prospect of exponential quantum speedup using tunneling.

Our technical contributions include extending the canonical path method for analyzing Markov chains to cover the case when not all vertices can be connected by low-congestion paths.  We also develop methods for taking advantage of warm starts and for relating the quantum state in QA to the probability distribution in SQA.  These techniques may be of use in future studies of SQA or of rapidly mixing Markov chains in general.}

\section{Introduction}

Classical algorithms are often useful but not provably so, with justifications for 
their success coming from a combination of empirical and heuristic
evidence.  For example, the simplex algorithm for linear programming was
successful for decades before being proven to run in polynomial time,
and for a long time was the most practical LP solver even while the
ellipsoid algorithm was the only provably poly-time solver.  Another
example is MCMC (Markov chain Monte Carlo) which is used for applications in statistics,
simulation, optimization and elsewhere, but almost never in regimes
that are covered by formal proofs of correctness.

With quantum algorithms, there has been necessarily a greater emphasis
on provable correctness.  The present state of quantum computing technology does not yet allow us to test 
large-scale quantum algorithms empirically, nor can we usually empirically determine whether 
a proposed quantum algorithm outperforms all classical algorithms on worst-case 
inputs.
Nevertheless, heuristic quantum algorithms are likely to be important for practical problems, just as they have been throughout the history of classical computing.

A particularly compelling heuristic proposal for optimization problems is quantum 
annealing (QA), also known as quantum adiabatic optimization~\cite{Nishimori:98a,farhi-2000}.  
(In this work we use the term ``quantum annealing'' to mean adiabatic optimization in thermal equilibrium at a low but non-zero temperature, though in some other contexts QA may be taken to include non-equilibrium thermal effects.)
The idea of QA is to
interpolate between a static problem-independent Hamiltonian such as
$-\sum_i \sigma_x^i$ for which we can efficiently prepare the ground
state, and a final Hamiltonian whose ground state yields the desired
answer.  If we want to minimize a function $f:\{0,1\}^n \ra \bbR$ then
we can take this final Hamiltonian to be proportional to $\diag(f)$.
This can be thought of as a quantum version of classical simulated annealing (SA) with the diagonal terms playing the role of bias and the
off-diagonal terms causing hopping.  Like SA its
performance is hard to make provable general statements about, but it
is a promising general-purpose heuristic, and rigorous statements about its performance are known for many illustrative cases.

Intriguingly, QA has been shown to have an exponential asymptotic advantage over simulated annealing for certain cost functions~\cite{farhi-2002}.  Two
examples are given in \cite{farhi-2002}: one in which the quantum
algorithm could be said to be taking advantage of symmetry (``the bush
of implications''), and another which models tunneling (``the spike'') that will be our primary focus here.  The cost function for the spike is,
\begin{equation}
   f(z) := \left\{
     \begin{array}{lr}
      |z| +n^{\alpha}  \; : \; n/4 -n^\zeta/2  < |z| < n/4 + n^\zeta/2\\
      |z|  \; \; \; \; \; \;  \; \; \; \; : o.w.
     \end{array},
   \right. \label{eq:spikedef}
\end{equation}
where $|z|$ is the Hamming weight of the string $z$, and $\alpha > 0$, $\zeta \geq 0$ are independent of $n$.  The global minimum of $f$ is the string with $|z| = 0$, but the spike term creates a local minimum at $|z| = n/4 + n^\zeta/2$.  The spike presents a problem for a simulated annealing algorithm which only proposes moves that flip $k$ bits for $k \leq n^\zeta$.   First recall the definition of SA: starting with a point $x\in \{0,1\}^n$, it repeatedly chooses a random nearby point $y$ (say within the Hamming ball of radius $k)$ and moves to $y$ with probability $\min(1, e^{(f(x)-f(y))/T})$, where $T$ is a temperature parameter that is gradually lowered.
Following \cite{farhi-2002}, consider first a SA algorithm that flips one bit at a time, i.e.~with $k=1$.  Since SA begins at high temperature the initial state is overwhelmingly likely to have Hamming weight near $n/2$, and as the temperature of the system is lowered the random walk will move to strings of lower Hamming weight until reaching the local minimum at $n/4 + n^\zeta/2$.  This will happen for $T=O(1)$, so at this point the probability of accepting a move onto the spike with probability $e^{-\Omega(n^\zeta)}$, so classical SA requires exponential time to find the global minimum with high probability.   This argument applies to flipping any $k < n^\zeta$ bits at once.  Now suppose that the SA algorithm flips $n^c$ bits at a time for some $c\leq 1$.  Once the Hamming weight is $\approx n/4$, flipping $n^c$ random bits will change the Hamming weight by a random variable with expectation $\approx \frac{1}{2}n^c$ and standard deviation $\approx \frac{\sqrt{3}}{2}n^{c/2}$.  This has probability $e^{-\Omega(n^c)}$ probability of being negative, so with high probability the SA algorithm will not even attempt to move past the spike. 
In contrast, for any $\alpha + \zeta < 1/2$ it can be shown that QA finds the global minimum with high probability in time $\cO(n)$~\cite{Reichardt-2004}, showing that an exponential separation in the perfomance of SA and QA is possible.  

While the spike is clearly a toy problem and can be solved efficiently by classical algorithms that exploit
its structure, an important aspect of both QA and SA is that a single, general implementation of these algorithms is meant to be useful for solving a large variety of different problems without knowledge of their structure.  Moreover, the spike arguably demonstrates a general advantage of QA over SA in tunneling through thin, high barriers in the energy landscape.  

On the other hand, the standard formulation of QA uses a stoquastic
Hamiltonian (i.e. a local Hamiltonian with non-positive off-diagonal matrix elements in the computational basis), and computational models based on ground states or thermal states of such systems are believed to be less powerful than universal quantum computation.  In addition to complexity theoretic evidence~\cite{bravyi-2006,bravyi-2006b}, suggestive evidence for this belief is also provided by the quantum-to-classical mapping of Suzuki et al.~\cite{suzuki-1977, suzuki-1986}, which allows for properties of low-energy states of a stoquastic Hamiltonian to be estimated using classical Markov chain Monte Carlo methods.  These algorithms are known as Quantum Monte Carlo (QMC) methods, and despite the name, are algorithms for classical computers.  While QMC for stoquastic Hamiltonians is always a well-defined algorithm, its performance depends on the rate at which a Markov chain converges to its stationary distribution.  This can range from polynomial to exponential time, and few general conditions are known in which it is provably polynomial-time. A few cases where the simulation can be made provably efficient are adiabatic evolution with frustration-free stoquastic Hamiltonians with a unique ground state~\cite{bravyi-2008} and ferromagnetic transverse Ising models in a large range of temperatures~\cite{bravyi-2014}, but while these have some physics significance, they do not translate into nontrivial cost functions for QA.

When QMC is applied to QA Hamiltonians the result is an algorithm called simulated quantum annealing (SQA).  Although there are examples for which standard versions of SQA take exponentially longer than the quantum evolution being simulated~\cite{hastings-2013}, the general challenge from SQA to QA remains: for any purported speedup of QA we should see whether it can also be achieved by SQA.   Moreover, since SQA is a Markov chain based algorithm on a domain that can be interpreted as a classical spin system, and since SQA is designed to sample from the output of a quantum optimization procedure, SQA can be considered as yet-another physics-inspired classical optimization method in its own right, which can naturally be compared to SA.  

The main result of this paper is that the standard version of SQA, which does not use any structure of the problem, finds the minimum of the cost function \eq{spikedef} in polynomial-time when $\alpha  + \zeta < 1/2$.   
\begin{theo} \label{thm:main-result}
Simulated quantum annealing based on the path-integral Monte Carlo method efficiently samples the output distribution of QA for the spike cost function \eq{spikedef} when $\alpha + \zeta < 1/2$.  The running time using single qubit worldline updates is $\tilde \cO(n^7)$, and the running time using single-site spin flips is $\tilde \cO(n^{17})$.    (Worldline and single-site spin flips are defined in \secref{sqaPrelim}.)
\end{theo}
Thus SQA obtains an exponential speedup over SA for this particular problem.  This result suggests that the benefit of adiabatic evolution in tunneling through barriers should not be thought of as an exclusively quantum advantage, since it can also be achieved by a general-purpose classical optimization algorithm.
We remark that the $\tilde\cO(n^{17})$ scaling is almost certainly  an artifact of our analysis and that numerical evidence suggests that $O(n^4)$ single-site updates should be sufficient~\cite{Crosson-2014b}.

\begin{table}[h]
\begin{center}
\begin{tabular}{cp{2cm}c}
\toprule
acronym & definition &  notes \\
\midrule
SA & simulated annealing & 
{Classical MCMC algorithm that simulates thermal annealing~\cite{annealing83}}
\\[7mm] QA & quantum annealing & 
{Adiabatic optimization~\cite{farhi-2000} generalized to low temperatures} \\[7mm]
SQA & simulated quantum annealing & 
{Classical MCMC that simulates quantum annealing; cf.~\secref{sqaPrelim}}
 \\
\bottomrule
\end{tabular}
\caption{Our paper repeatedly uses the similar sounding acronyms SA, QA and SQA, which are defined here.  QA is a quantum algorithm while SA and SQA run on classical computers.
\label{tab:acronyms}}
\end{center}
\end{table}

\subsection*{Previous Work}

There have been many past studies comparing the performance of SA and SQA using numerics~\cite{Martonak-2002, Santoro-2005, Inack-2015} and more recently using analytical methods of physics such as the instanton approximation to tunneling~\cite{Isakov-2015}.  Studies comparing QA to SQA have also begun to emerge since~\cite{Boixo-2014} found the success probabilities of SQA are highly correlated with the results of QA performed on D-Wave quantum hardware with hundreds of qubits, while the distribution of success probabilities for SA on the same set of instances bears little resemblance to that of QA and SQA.  More recently, the performance of QA, SQA, and SA was empirically compared on an ensemble of spin glass instances with were designed to have tall, thin barriers~\cite{Denchev-2015}, as a step towards understanding the kinds of instances for which QA has an advantage over SA.   In that work QA and SQA were found to have roughly the same scaling with system size for that particular ensemble of instances, though it was also pointed out that the large constant overhead in SQA made it less competitive in the sense of wall-clock times using modern classical hardware.   

Without access to quantum hardware, comparison of SQA and QA is either limited to small system sizes where QA Hamiltonians can be exactly diagonalized ($\lesssim 50$ qubits), or to models for which analytical solutions of the quantum system are known (such as the spike problem we study here).  We remark that the spike and related objective functions have the subject of recent analytic work~\cite{Kong-2015, Brady-2016}, and that there have also been numerical studies of SQA~\cite{Crosson-2014b,Brady-2015}, with findings that are consistent with our main result.

\subsection*{Proof Outline}

Our proof of the efficient convergence of SQA on the spike problem involves bounding the
mixing time of the underlying Markov chain, and there is an interesting parallel
between a method which was used to lower bound
the QA spectral gap when $\alpha +\zeta < 1/2$~\cite{Reichardt-2004}. 
There, a lower bound on the quantum gap can be found using a variational method with a trial wave function equal to the ground state of the system when no spike term is present (i.e. QA for the spikeless Hamming weight cost function $\tilde{f}(z) = |z|$).  Similarly, we compare the spectral gap $\lambda$ of the SQA Markov chain for the spike system with the spectral gap $\tilde{\lambda}$ of the spikeless system (throughout the subsequent sections we use tildes to distinguish quantities belonging to the spikeless system).  Without a spike term, the quantum Hamiltonian $\tilde{H}$ is a tensor product operator with no interactions between the qubits.  This trivial system translates in SQA to a collection of $n$ non-interacting 1D classical ferromagnetic Ising models in a uniform magnetic field (which will become clear when the SQA Markov chain is described in detail in \secref{sqaPrelim}), and upper bounding the mixing time for this system is relatively straightforward.
 
Let $\pi$ and $\tilde{\pi}$ be the stationary distribution of the SQA Markov chain with and without the spike.  These stationary distributions are close in a sense, $\|\pi - \tilde{\pi}\|_1 < \poly(n^{-1})$, but on the other hand there are exponentially many points $x \in \Omega$ for which the ratio $\pi(x)/\tilde{\pi}(x)$ is exponentially small.  A review of existing comparison techniques in \secref{comparison} concludes that none is quite suited to the present problem; indeed the review \cite{Dyer-2004} states that there have been ``relatively few successes in comparing chains with very different stationary distributions''.   To overcome this we introduce a comparison method which involves partitioning the state space into ``good'' and ``bad''  sets of vertices, $\Omega = \Omega_G \cup \Omega_B$.  In \secref{almost-canonical} we begin with a set of canonical paths yielding a bound $\tilde{\rho}$ on the congestion of the easy-to-analyze chain, and show that the paths which lie entirely within $\Omega_G$ can be used to construct an upper bound on the congestion $\rho$ of the difficult-to-analyze chain, albeit within the set $\Omega_G$ of measure less than 1.  This ``most-paths comparison'' method may be of independent interest, and so the exposition in \secref{almost-canonical} is given without dependence on the specific details of SQA.  

There are two main ingredients to the comparison method.  First we show that if two chains have stationary distributions and transition probabilities that are similar on most of the points, then we can convert a known gap for one chain into a set of canonical paths for \emph{most} of the state space of the other.

\begin{theo}[Most-paths comparison]\label{thm:most-compare-inf}
Let $(\pi,P)$ and $(\tilde{\pi},\tilde{P})$ be reversible Markov chains with the same state space graph $(\Omega,E)$.  Let $a = \max_{x\in \Omega} \pi(x)/\tilde\pi(x)$ and define $\Omega_\theta := \{x\in \Omega : \pi(x) < \theta\tilde\pi(x)\}$.  If there is a set of canonical paths for $(\tilde \pi, \tilde P)$ achieving congestion $\tilde \rho$ and satisfying $3 a^2 \tilde \rho \tilde \pi(\Omega_\theta) < 1$, then there is a subset $\Omega_G\subset \Omega$ with $\pi(\Omega_G) \geq 1 - 3 a^2 \tilde \rho \, \theta \pi(\Omega_\theta)$, and a canonical flow for $(\pi,P)$ that connects every $x,y \in \Omega_G$ with paths contained in $\Omega_G$ for which the congestion $\rho$ of any edge in $\Omega_G$ satisfies
\be
\rho \leq 16 \; \theta \max_{x,y\in \Omega_G} \left [\frac{\tilde{P}(x,y)}{P(x,y)} \right ] a^2 \tilde \rho.
\ee
\end{theo}

There is a caveat here, which is that this bound on the congestion applies to the transitions of the Markov chain $P$ on the subset $\Omega_G$.  If we assume that walkers leaving $\Omega_G$ are deleted then, since $P$ is not restricted to $\Omega_G$, these transitions form a substochastic ``leaky'' random walk on the set $\Omega_G$, with a quasi-stationary distribution equal to $\pi$ within this subset.  (The term ``quasi-stationary'' refers to the fact that in the infinite time limit repeated applications of $P|_{\Omega_G}$ will converge to zero, but there may be a long intermediate time when we are close to $\pi$.)

Thus it is necessary to show that the chain mixes before it leaves the good set $\Omega_G$.  
One way to guarantee this is to use a ``warm start,'' meaning a starting sample from a distribution that is close to the quasi-stationary distribution.   Our analysis of SQA will rely on following the adiabatic path used by the quantum algorithm to guarantee the warm-start condition.  

\begin{theo}\label{thm:leaky-convergence-informal}
Let $(\pi, \Omega, P)$ be a reversible Markov chain and suppose $\Omega = \Omega_G \cup \Omega_B$ is a partition.  Let $P_G$ be the substochastic transition matrix $P_G(x,y) := P(x,y) 1_{x\in \Omega_G} 1_{y\in \Omega_G}$.  Suppose there is a set of canonical paths connecting every pair of points $x,y\in \Omega_G$, and the congestion of the walk $P$ on this set of paths is $\rho$.  If $\mu$ is a warm start with $\mu(x) \leq M \pi(x)$ for all $x\in \Omega_G$ then the distribution obtained by starting from $\mu$ and applying $t$ steps of the random walk satisfies
\be
\| \mu P_G^t - \pi \|_1  \leq Mt\pi(\Omega_B) + \pi_{\min}^{-1} e^{-  t/\rho} \label{eq:leaky-convergence-informal}
\ee
\end{theo}

The SQA state space can be interpreted as a path (worldline) representation of the original quantum system, and the bad states which constitute $\Omega_B$ will be those for which the paths spend too much ``time'' on the location of the spike (i.e. on strings with Hamming weight between $n/4 - n^\zeta/2$ and $n/4 + n^\zeta/2$).  States that spend too much time on the spike are those for which $\pi(x)/\tilde{\pi}(x)$ is exponentially small, and naturally those are the ones we will need to exclude.   In \secref{spike-time} we show that the mean spike time is proportional to the square of the ground state amplitude on the spike, while the $m$-th moment of the spike time distribution can also be bounded using the properties of the corresponding quantum system.   Finally, we use the derived upper bound on the $m$-th moment of the spike time distribution to upper bound the probability of large deviations from the mean spike time, which yields an upper bound on $\pi(\Omega_B$) that suffices to complete the proof.

\subsection*{Discussion}\label{sec:discussion}
Our proof does not bound the convergence time for SQA $\alpha + \zeta >1/2$, although QA does work for some values of $(\alpha, \zeta)$ in this range, such as when $\zeta=0,\alpha=O(1)$~\cite{Kong-2015} or $\alpha + 2\zeta<1$~\cite{Brady-2016}.  We conjecture that SQA will be efficient for these values as well (which is supported by numerical evidence), though this will require extensions of the present techniques.  The approach used in this work is also suggestive of a more general connection between the quantum spectral gap of Hamming symmetric barrier problems (including barriers of various shapes and widths) and the corresponding performance of SQA, which we sketch here.

Assume we are near the critical value $s^*$ of the adiabatic parameter at which the system tunnels through the barrier i.e. for $s < s^*$ the ground state probability mass is concentrated on one side of the barrier, while for $s > s^*$ the opposite occurs, and for $s = s^*$ the probability mass on both sides of the barrier is $\cO(1))$.  While a general understanding of what barriers admit such a tunneling description has not yet been found, there are many examples for which this is known to occur\cite{farhi-2002, Kong-2015, Lidar-2015b}.   Assume that the QA spectral gap at $s^*$ is $\Delta = 1/\poly(n)$, so that QA will be able to pass through the barrier efficiently.  

We expect the first excited state to have a node at some location $|k\rangle$ inside the barrier region, and some properties of the ground state wave function near $|k\rangle$ can be inferred from the spectral gap.  Since the spectral gap is at least $1/\poly(n)$, we know that the ground state wave function cannot be too small inside of the barrier, or else there would be a balanced-weight cut in the ground state that could be used to construct an orthogonal state with low energy.  This benefits the comparison approach because the ground state amplitudes inside the barrier need to be at least $1/\poly(n)$ in order for a $1/\poly(n)$ fraction of the canonical paths to transfer from the spikeless system to the system with a barrier.  

On the other hand, the amplitudes inside the barrier cannot become too large because the barrier is a classically forbidden region, and because an upper bound on $\Delta$ will also upper bound the amplitudes near $|k\rangle$ (by the same argument using a balanced-weight cut together with the assumption that the first excited state has a node at $|k\rangle$).  The fact that the total probability mass inside the barrier is not too large could be useful for deriving the necessary upper bounds on $\pi(\Omega_\theta)$ that are used in the comparison approach.  The primary ingredient that would be needed to make this sketch rigorous is a better  understanding of the wave functions and spectral gap of the quantum system with a barrier, and depending on the outcome of this understanding the comparison approach for analyzing SQA may require modifications as well.  

More generally, we believe that our most-paths comparison methods should have wider applicability.  For example, consider a collection of classical particles with weak repulsive interactions.  If the particles were non-interacting the thermal distribution of the particles would be easy to sample from, and if the interactions are weak enough, then they do not shift the typical probabilities (or energies) by very much.  While some configurations will have exponentially lower probability in the interacting case (if many particles are very close to each other), these configurations should be overall very unlikely.  In this setting our framework should imply that
the repulsive interactions do not significantly worsen the mixing time.

Finally, while relatively few rigorous facts are known about the performance of SQA or Quantum Monte Carlo more generally, it remains in practice a successful and widely used class of algorithms.  This strikes us as an area where theorists should work to
catch up with current practice.

\subsection*{Overview of remaining sections}
In \secref{background} we review the QA and SQA algorithms in the forms that our paper will use them.  \secref{incomplete} fleshes out our Theorems \ref{thm:most-compare-inf} and \ref{thm:leaky-convergence-informal} on incomplete sets of canonical paths and leaky random walks. This section presents its results for a general Markov chains and can be understood without reference to the specific SQA Markov chains discussed elsewhere in the paper.  Our main result is proved in \secref{spike-time}; this entails showing that the Markov chains from \secref{background} meet the mixing conditions laid out in \secref{incomplete}.

\section{Background}\label{sec:background}
\subsection{Quantum annealing}
Quantum annealing associates a cost function $f:\{0,1\}^n \rightarrow \mathbb{R}$ with a Hamiltonian that is diagonal in the computational basis,
\be
H_f := \sum_{z \in \{0,1\}^n} f(z) |z\rangle \langle z| \quad , \quad  
\ee
so that the ground state of $H_f$ is a computational basis state corresponding to the bit string that minimizes $f$.  To prepare the ground state of $H_f$ the system is initialized in the ground state of a uniform transverse field, which can be easily prepared, 
\be
 H_0 := -\sum_{i = 1}^n  \sigma^x_i \quad , 
\quad  |\psi_{\textrm{init}}\rangle := \frac{1}{\sqrt{2^n}} \sum_{z\in \{0,1\}} |z\rangle ,
\ee
and then linearly interpolates between $H_0$ and $H_f$,  
\begin{equation}
H := H(s) = \left(1-s\right)H_0 + s H_f \; \; ,  \label{eq:aitchess}
\end{equation}
where the adiabatic parameter $s$ sweeps through the interval $0 \leq s \leq 1$.   The total run time $t_{\max}$ of the algorithm depends on how quickly the adiabatic parameter is adjusted, which defines a time-dependent Hamiltonian $H(t) := H(s = t/T)$.  At zero temperature the system evolves according to the Schr\"odinger equation, $ \frac{d}{dt}|\psi(t)\rangle = -iH(t) \psi(t)$, and the adiabatic theorem ensures that the state $\psi(T)$ at the end of the evolution has a high overlap with the ground state of $H_f$ as long as $T\geq \textrm{poly}(n,\Delta^{-1})$, where $\Delta = \min_s E_1 (s) - E_0(s)$ is the minimum gap between the two lowest eigenvalues of $H(s)$ during the evolution.  

More generally (and realistically) we can take the state of the system to be not the ground state but a thermal state with inverse temperature $\beta<\infty$.  
The equilibrium thermal state of the system evolves with the adiabatic parameter,
\be
\sigma(s) := \frac{e^{-\beta H(s)}}{\cZ(s)} \quad , \quad \cZ(s) := \tr e^{-\beta H(s)}.  \label{eq:thermalQA}
\ee
We can see that if $\beta=\infty$ the system will be in the ground state, and we will assume that $\beta$ is sufficiently large so that the system will remain close to the ground state.  Just as adiabatic theorem has an error term corresponding to transitions out of the ground state, in thermal annealing the system will in general not be exactly in the equilibrium state \eq{thermalQA}.   Nevertheless it is a useful idealization, and one that our simulation algorithms will aim to reproduce.
Specifically the simulated quantum annealing algorithm described in the next section will produce samples from the distribution $\Pi_s(z) :=  \langle z | \sigma(s) | z \rangle$.   The minimum gap $\Delta$ of the Hamiltonian \eq{aitchess} with the cost function \eq{spikedef} is constant when $\alpha + \zeta < 1/2$~\cite{Reichardt-2004}, and the density of states is such that taking $\beta = n^{\epsilon}$ for a constant $\epsilon > 0$ suffices to make $\left\| \;\rho(s) - |\psi_0(s)\rangle\langle \psi_0(s)|\; \right\|_1 < \cO(n e^{-n^\epsilon})$.  Thus this low-temperature thermal-equilibrium version of QA produces a final state which can be sampled to obtain the minimum of $f$.  

\subsection{Simulated quantum annealing}\label{sec:sqaPrelim}
In principle stoquastic Hamiltonians such as \eq{aitchess} are amenable to a variety of classical Markov chain based simulation algorithms, which are collectively known as quantum Monte Carlo (QMC) methods (the term ``stoquastic" is a combination of ``quantum'' + ``stochastic'' in the sense of stochastic matrices~\cite{bravyi-2006}).   Any QMC method applied to the QA Hamiltonian \eq{aitchess} defines a version of SQA.  The version we consider here is based on the path-integral representation of the thermal state \eq{thermalQA}.

The starting point of the method is to express $\cZ(s)$ as a sum over an exponential number of nonnegative terms, or in physics language to write the quantum partition function as an imaginary-time path integral over trajectories basis states.  Since the Hamiltonian is stoquastic in the standard basis, these trajectories will have the form $(x_1,...,x_L)$, where $x_i \in \{0,1\}^n$,
\begin{equation}
\mathcal{Z} = \tr e^{-\beta H} = \tr \prod_{i = 1}^L e^{-\frac{\beta H}{L}} = \sum_{x_1,...,x_L} \prod_{i=1}^L \langle x_i | e^{-\frac{\beta H}{L}} | x_{i+1} \rangle, \label{eq:expandPartition}
\end{equation}
where $x_{L+1} := x_1$, $L$ is to be chosen in the next step and we neglect the dependence on $s$ to keep the notation simple.  The individual bit strings $x_i$ in $(x_1,...,x_L)$ are sometimes called``time slices.'' When $\beta \| H\|/L$ is sufficiently small the Suzuki-Trotter approximation provides a way to split up the non-commuting terms in $e^{-\beta H/ L}$ while only incurring a small error in the partition function.  Define $A = -\beta s H_f$ and $B = -\beta (1-s) H_0$, so that the partition function is $\mathcal{Z} = \tr e^{A + B}$.   Define the Suzuki-Trotter approximation to the partition function,
\begin{equation}
Z := \tr \left [ e^{\frac{A}{L}}e^{\frac{B}{L}} \right ]^L . \label{eq:ST}
\end{equation}
According to Lemma 3 and the surrounding discussion of \cite{bravyi-2014}, taking $L = \Theta\left(\sqrt{(\beta \|H\|)^3\delta^{-1}}\right)$ for sufficiently small $\delta > 0$ achieves
\be
(1-\delta) \mathcal{Z} \leq Z \leq \mathcal{Z}(1 + \delta). \label{eq:suzukiTrotter}
\ee
We will take $\delta = n^{-1}$, which implies $L = \cO(n^{2}\beta^{3/2})$ and subsequently ignore the error in \eq{suzukiTrotter} because it does not affect the convergence time of SQA, and creates only a negligible error in the distribution which SQA will sample from for the spike cost function \eq{spikedef}.  Expanding \eqref{eq:ST} as was done for \eqref{eq:expandPartition},
\begin{align}
Z &= \sum_{x_1,...,x_L} \prod_{i=1}^L \langle x_i | e^{\frac{A}{L}}e^{\frac{B}{L}} | x_{i+1} \rangle\\
&= \sum_{x_1,...,x_L}  e^{-\frac{\beta s}{L}\sum_{i=1}^L f(x_i)}\prod_{k=1}^L \langle x_k | e^{\frac{B}{L}} | x_{k+1} \rangle\\
&= \sum_{x_1,...,x_L}  e^{-\frac{\beta s}{L}\sum_{i=1}^L f(x_i)}\prod_{j=1}^n \prod_{k=1}^L \langle x_{j,k} | e^{\omega \sigma^x_j} | x_{j,k+1} \rangle  , \label{eq:QtCmapping3}
\end{align}
where $\omega := \beta(1-s)/L$.  Using the identity $\exp\left(\omega \sigma^x \right) =  \cosh(\omega) I + \sinh(\omega)\sigma^x $, the individual factors of the product in \eqref{eq:QtCmapping3} become
\begin{equation}
\langle x_{j,k} | e^{\omega\sigma^x_j} | x_{j,k+1} \rangle =  \cosh(\omega) \left [1_{x_{j,k} = x_{j,k+1}} + \tanh(\omega) 1_{x_{j,k} \neq x_{j,k+1}} \right ],
\end{equation}
and after suppressing the uninteresting multiplicative factor of $\cosh(\omega)^{n L}$, the partition function is expressed as
\begin{equation}
Z =  \sum_{x_1,...,x_L}  e^{-\frac{\beta s}{L}\sum_{i=1}^L f(x_i)}\prod_{j=1}^n \prod_{k=1}^L  \left [1_{x_{j,k} = x_{j,k+1}} + \tanh(\omega) 1_{x_{j,k} \neq x_{j,k+1}} \right ]\label{eq:QtCmapping4},
\end{equation}
and so $Z$ can be viewed as the normalizing constant of a probability distribution,  
\ba
\pi(x_1,...,x_L)  &= \frac{1}{Z} e^{-\frac{\beta s}{L} \sum_{i=1}^L f(x_i)} \prod_{j,k= 1}^{n,L}  \left [1_{x_{j,k} = x_{j,k+1}} + \tanh\left(\omega\right) 1_{x_{j,k} \neq x_{j,k+1}} \right ]\\
& = \frac{1}{Z} e^{-\frac{\beta s}{L} \sum_{i=1}^L f(x_i)} \prod_{j =1}^{n} \phi(\bar{x}_j) \label{eq:pimain}
\ea
where $\bar{x}_j := (x_{j,1}, ... , x_{j,L})$ is called ``the worldline of the j-th qubit", and $\phi(\bar{x}_j) := \tanh(\omega)^{\left|\left\{k : x_{j,k} \neq x_{j,k+1}\right\}\right|}$ counts the number of consecutive bits which disagree in that worldline.   

Performing a similar calculation for $\Pi(x) = \langle x | \sigma | x \rangle$ (where $\sigma=e^{-\beta H}/\cZ$) one finds it can be expressed as the marginal of $\pi$ on the first time slice,
\be
\Pi(x) = \sum_{x_2,...,x_L} \pi(x,x_2,...,x_L) \label{eq:capitalPi}
\ee

From this point SQA proceeds by discretizing the adiabatic path and using the Markov chain Monte Carlo method to sample from $\pi$ at various values of the adiabatic parameter $s_1\approx 0, \ldots ,s_{\max}\approx 1$.  We will analyze two discrete-time Markov chains which both have stationary distribution $\pi$ on the state space $\Omega = \{0,1\}^{n\times L}$.  The first chain consists of \textbf{single-site Metropolis updates}.  If $x,x' \in \Omega$ differ by a single bit, the transition probability from $x$ to $x'$ is
\begin{equation}
P_M(x,x') = \frac{1}{2nL}\min \left \{1, \frac{\pi(x')}{\pi(x)} \right \}, \label{eq:Metropolis}
\end{equation}
and otherwise the transition probability is zero.  We interpret this as follows: with probability 1/2 propose to flip one of the $nL$ bits in $x$ to create $x'$, and accept this move with probability $\min(1,\pi'(x)/\pi(x))$.  Otherwise the configuration $x$ remains unchanged.

Note that $\pi$ is supported on all of $\Omega$ when $s < 1$, while at $s = 1$ the state space becomes disconnected under the local move transitions described above.  This is not a major limitation for our application, however, since sampling from $\Pi$ when $s_{\max} = 1 - \frac{1}{n}$ suffices to find the true minimum of $f$.  In practical implementations it is important to avoid this ``critical slowing down'' as $s\rightarrow 1$ by using cluster updates that flip many spins at once.  Since local moves in classical SA correspond to flipping a single bit $\{1,\ldots,n\}$, one could argue that a move which arbitrarily updates the worldline of a single qubit should be considered a local move in SQA as well.  More importantly, such moves can be performed efficiently for any easily computable objective function $f$.  Therefore in addition to the single-site Metropolis updates defined above we will analyze the \textbf{single qubit heat-bath worldline updates}, which are a form of generalized heat-bath updates~\cite{dyer2002mixing}.

\begin{dfn*}
The heat-bath worldline update $(\bar{x}_1,\ldots\bar{x}_n) \rightarrow (\bar x_1',\ldots,\bar x_n')$ proceeds as follows: 

\begin{enumerate}
\item Select a site $i \in \{1,\ldots,n\}$ uniformly at random.
\item Set $\bar{x}'_j = \bar{x}_j$ for all $j\neq i$.
\item Choose $\bar{x}'_i$ from the conditional distribution $\pi(\bar{x}'_i |\bar{x}'_1,\ldots,\bar{x}'_{i-1},\bar{x}'_{i+1},\ldots,\bar{x}'_n)$.  
\end{enumerate}
\end{dfn*}
As with all generalized heat-bath updates these transitions define a Markov chain which is reversible with respect to $\pi$.  An algorithm that efficiently implements these transitions is given in \cite{farhi2009quantum} and without using any structure of the cost function \eq{spikedef} it runs in $\tilde \cO(n^2 \beta^2)$ elementary steps with high probability.   One practical advantage of these updates is that $L$ can be taken to be effectively $\infty$ (or more precisely, the runtime can be made to scale as $\poly\log(L)$).  This is because the typical number of bit flips per worldline is independent of $L$, and so we can represent $\bar x_i$ succinctly by specifying only the locations of the bit flips.   This is a major improvement in practice but for the purposes of establishing poly-time mixing, we will not explore this further here.

At each value of the adiabatic parameter, the run-time of SQA will be determined by the mixing time of either of the Markov chains described above.  This quantity can be defined in terms of the total variation distance from $\pi$ to the distribution $P^t(x,\cdot)$ obtained by running the chain for $t$ steps starting from $x$, 
\begin{equation}
d_{x}(t) := \max_{A \subseteq \Omega} |P^t(x,A) - \pi(A)| = \frac{1}{2} \sum_{x'\in \Omega}|P^t(x,x')-\pi(x')|,
\end{equation}
with the mixing time $\tau(\epsilon)$ being the worst-case time needed to be within variation distance $\epsilon$ of the stationary distribution,
\begin{equation}
t_{\mix}(\epsilon) := \max_{x\in \Omega} \min_t \{t: d_{x}(t') \leq \epsilon \; \; \forall t \geq t'\}.
\end{equation}
A standard way to bound the mixing time is to relate it to the spectral gap $\lambda$ of the transition matrix $P$~\cite{peres-2008}. For all $x \in \Omega$,
\be \|  P^t(x,\cdot)  - \pi \|_1 \leq \pi_{\min}^{-1} e^{-\lambda
  t}.
\label{eq:mixing-gap}\ee
which implies $t_{\mix}(\epsilon) \leq \lambda^{-1} \log\left(\frac{1}{\epsilon \pi_{\min}}\right)$.
These bounds are worst-case in the sense that they handle any starting vertex $x\in \Omega$.
In the next section of our paper, we will develop slightly different methods to deal with the fact that our Markov chain may not mix efficiently from some starting vertices.

\section{Incomplete sets of canonical paths}\label{sec:incomplete}

In this section we discuss how to show rapid mixing even in the
presence of a small number of bad vertices.  While we will freely make
assumptions specific to our particular problem we will introduce some
techniques that apply more generally to the analysis of Markov chains,
and may be of use elsewhere.  The notion of ``good'' and ``bad'' sets
in Markov chains has been used before~\cite{JRS04,CLMST14}, but always
(to our knowledge) in a setting where a separate argument shows the
bad set can always be quickly escaped.  By contrast we model the bad
set quite pessimistically and can assume that the walker gets absorbed
(or equivalently, trapped for an exponential amount of time) upon
hitting a bad vertex.  Despite this we will show that the overall algorithm
works with high probability.

\subsection{Markov chains and the path comparison method}\label{sec:comparison}
This subsection reviews some standard facts from section 13.5 of the book by Levin, Peres and
Wilmer~\cite{peres-2008}.  Let $(\pi,P)$,$(\tilde \pi, \tilde P)$ be reversible Markov chains and define $Q(x,y) := \pi(x) P(x,y) $ and
likewise for $\tilde{Q}$.  Define the inner product $\langle
f,g\rangle_\pi := \sum_x \pi(x) f(x)g(x)$ and the Dirichlet form 
\be \cE(f,g) := \langle \cE f, g\rangle_\pi := \langle (I-P) f, g\rangle_\pi.\ee
Lemma 13.11 of \cite{peres-2008} states that
\be \cE(f) := \cE(f,f) = \frac{1}{2} \sum_{x,y\in \Omega}
[f(x)-f(y)]^2 Q(x,y).\ee
This can be used to define the gap (cf.~Remark 13.13 of \cite{peres-2008}),
\be \lambda := \min_{\substack{f\in \bbR^\Omega \\ f\perp_\pi
    \mathbf{1}, \|f\|_2=1}} \cE(f) = 
\min_{\substack{f\in \bbR^\Omega \\ \Var_\pi(f)\neq 0}}
\frac{\cE(f)}{\Var_\pi(f)}.\label{eq:gap-def}\ee

To estimate the gap we will use various comparison methods.  Given
$x,y\in \Omega$ let $\cP_{xy}$ be the set of simple paths from $x$ to
$y$, and suppose that $\nu_{xy}$ is a measure over this set.  Then we
can define a congestion ratio:
\ba 
\rho(e) & := \frac{1}{Q(e)} \sum_{x,y\in \Omega} \tilde Q(x,y)
\sum_{\Gamma :  e \in \Gamma \in \cP_{xy}} \nu_{xy}(\Gamma) |\Gamma|
\label{eq:rho-e-def}\\
\rho & := \max_{e\in E} \rho(e) \label{eq:rho-def}
\ea
This last sum ranges over all paths $\Gamma \in \cP_{xy}$ that contain the edge $e$ and thus is a measure of the total load on edge $e$.
Then Corollary 13.26 of \cite{peres-2008} states that
\be \tilde\lambda \leq \left[ \max_{x\in \Omega}
  \frac{\pi(x)}{\tilde\pi(x)}\right]
\rho \lambda.\label{eq:mc-comparison}\ee

If $\nu_{xy}$ has all its weight on a single flow $\gamma_{xy}$ then
we can slightly simplify \eq{rho-e-def} to
\be \rho(e) = \frac{1}{Q(e)} \sum_{\substack{x,y\in \Omega \\ e \in
    \gamma_{xy}}}
 \tilde Q(x,y) |\gamma_{xy}|.
\label{eq:single-path}\ee
Another variant is when we compare a chain $\tilde Q$ with the
complete graph for which the $x\ra y$ transition probability is simply
$\tilde \pi(y)$.  The paths used for this purpose are labelled $\{\tilde\gamma_{xy}\}$.
The corresponding congestion is 
\be \tilde\rho(e) := 
\frac{1}{\tilde Q(e)} \sum_{\substack{x,y\in \Omega \\ e \in
    \tilde \gamma_{xy}}}
 \tilde \pi(x)  \tilde \pi(y) |\tilde \gamma_{xy}|.
\label{eq:compare-complete}\ee
Since the complete graph has gap 1 and the same stationary
distribution, we have
\be \tilde \lambda \geq \frac{1}{\tilde\rho}.\ee

\subsection{Most-paths comparison}\label{sec:almost-canonical}
In this section we will describe a comparison method for two reversible Markov chains $(\tilde \pi, \tilde P)$, $(\pi, P)$ defined on the same state space graph $(\Omega,E)$.  The method is designed for chains which are comparable on a subset $\Omega_G$ containing most of the stationary probability of both chains, but which may have $\pi$ and $\tilde \pi$ differ greatly in other regions of the state space.  Assuming there is a set of canonical paths connecting all $x,y \in \Omega$ with congestion $\tilde \rho$ for $(\tilde \pi, \tilde P)$, then we will show it is possible to construct a canonical flow for $(\pi,P)$ connecting all $x,y\in\Omega_G$ which has comparable congestion $\rho$ within $\Omega_G$.   

In our application of this method the easy-to-analyze chain $(\tilde \pi, \tilde P)$ will be the SQA Markov chain for the spikeless distribution, and $(\pi, P)$ will be the SQA chain for the spike system.  Therefore $\pi(x)$ will never be much larger than $\tilde \pi(x)$, and the quantity
\be
a := \max_{x\in \Omega} \frac{\pi(x)}{\tilde \pi(x)}\label{eq:pi-ratio}
\ee
will be $\cO(1)$ for our application.  However, there will be exponentially many configurations $x\in\Omega$ with $\pi(x) / \tilde \pi(x) = \cO(e^{-n})$ and so we define a subset parameterized by the value of this ratio,
\be
\Omega_\theta := \left \{x \in \Omega :  \frac{\pi(x)}{\tilde \pi(x)} < \theta \right\}.
\ee
Let $\{\tilde\gamma_{xy}\}$ be the set of paths for $(\tilde \pi, \tilde P)$ for which the congestion (defined in
\eq{compare-complete}) is $\tilde\rho$.  These paths may lead to far
larger congestion for $(\pi,P)$ if they involve routing flow over
edges $e$ where $ Q(e) \ll \tilde Q(e)$.  However, our first claim is
that most of these paths should avoid the bad set $\Omega_\theta$.
First observe that for any $e$ the definition of $\tilde\rho$ implies
that
\be  \sum_{\substack{x,y : \gamma_{xy}\ni e }} \tilde\pi(x) \tilde \pi(y)
\leq \tilde\rho \tilde Q(e). \label{eq:canonical-again}\ee
Define $E_\theta$ to be the set of edges incident upon
$\Omega_\theta$.  Reversibility implies that
 \be 
\frac 1 2\tilde\pi(\Omega_\theta) \leq \tilde Q(E_\theta) \leq
\tilde\pi(\Omega_\theta).\ee
(The inequalities are because an edge may be counted once or twice
depending on whether both end points are in $\Omega_\theta$.)  
Additionally for $e=(v,w)\not\in E_\theta$ we have
\be \tilde Q(e) = \tilde\pi(v) \tilde P(v,w) 
\leq \theta\pi(v) \tilde P(v,w)
\leq \theta R \pi(v) P(v,w) =  \theta R Q(e),
\label{eq:Q-comparison}\ee
where $R := \max_{(v,w) \notin E_\theta} \tilde{P}(v,w)/P(v,w)$.  Note that in our application $R$ will be $\cO(1)$ for all the types of transitions we consider.

Next we  Let $C := \Omega^2$, $C_B := \{(x,y) \in C :  \gamma_{xy}\cap
E_\theta \neq \emptyset\}$ and $C_G := C - C_B$.
  Now we sum \eq{canonical-again} over all
$e\in E_\theta$ to obtain
\be  \sum_{(x,y) \in C_B} \tilde\pi(x) \tilde \pi(y) |\tilde \gamma_{xy}|
\leq \tilde\rho \tilde Q(E_\theta)
=  \tilde\rho \tilde \pi(\Omega_\theta). \label{eq:not-so-bad}\ee
We conclude that not many of the $\tilde\gamma_{xy}$ go through any
edges in $E_\theta$.

Now we define a second partition of $\Omega$ into good and bad
vertices.  Let $C_B(x) := \{y : (x,y) \in C_B\}$ and define $C_G(x)$
similarly.  Then define 
\be \Omega_B := \{ x : \tilde\pi(C_B(x)) \geq 1/3a \},\ee
and $\Omega_G = \Omega-  \Omega_B$.
From \eq{not-so-bad} we have $\tilde\pi(\Omega_B) \leq 3a \tilde\rho
\tilde \pi(\Omega_\theta)$.


Using first \eq{pi-ratio} and the assumption that $\pi(\Omega_\theta)$
is sufficiently small we have
\be \pi(\Omega_G) = 1 - \pi(\Omega_B) \geq 1-3a^2\tilde\rho\tilde
\pi(\Omega_\theta) \geq \frac{11}{12}.  \label{eq:ThetaToBee}\ee
Thus $\Omega_G$ is a large-measure set that is mostly well connected.  Indeed $\forall x\in
\Omega_G$, $\pi(C_G(x))\geq 2/3$.

We can now define canonical flows between all pairs $x,y\in \Omega_G$.
Observe that 
\be \pi(C_G(x)\cap C_G(y)\cap \Omega_G) \geq 1/4.\ee
This means that even if $x,y$ are not directly connected to each other
by a path that avoids $E_\theta$, they are still indirectly connected
via pairs of paths that route through a large-measure $(\geq 1/4)$
set.  We will see that this allows us to construct canonical flows
that are not too much worse than our original canonical paths.

Observe next that the conditional
distribution $\pi_{xy} := \pi|_{C_G(x)\cap C_G(y) \cap \Omega_G}$ satisfies
$\pi_{xy} \leq 4 \pi$.  We will construct our flow $\nu_{xy}$ by
choosing a random $r \sim \pi_{xy}$ and concatenating the paths
$\tilde\gamma_{xr}$ and $\tilde\gamma_{ry}$.  The load on edge $e$
from these flows is 0 if $e\not\in E_\theta$, or if $e\in E_\theta$
can be bounded as
\ba
\sum_{x,y\in \Omega_G} \pi(x) \pi(y) & \sum_r \pi_{xy}(r) 
(1_{e\in\tilde\gamma_{xr}}+1_{e\in\tilde\gamma_{ry}})
(|\tilde\gamma_{xr}|+|\tilde\gamma_{ry}|)
\\ & \leq 4 \sum_{x,y\in \Omega_G} \pi(x) \pi(y) \sum_r \pi_{xy}(r) 
1_{e\in\tilde\gamma_{xr}}|\tilde\gamma_{xr}| & \text{by symmetry} \\
& \leq  16 \sum_{x,y\in \Omega_G} \pi(x) \pi(y) 
\sum_{r \in C_G(x)\cap \Omega_G} \pi(r) 
1_{e\in\tilde\gamma_{xr}}|\tilde\gamma_{xr}| & \text{since $\pi_{xy} \leq 4\pi$ }\\
& = 16 \sum_{x\in \Omega_G} \pi(x) \sum_{r \in C_G(x)\cap \Omega_G} \pi(r)
1_{e\in\tilde\gamma_{xr}} |\tilde\gamma_{xr}|& \text{since $\pi$ is normalized}\\
& \leq  16 \sum_{(x,r)\in C_G} \pi(x)\pi(r)
1_{e\in\tilde\gamma_{xr}} |\tilde\gamma_{xr}|& \text{passing to a superset}\\
& \leq  16a^2 \sum_{(x,r)\in C_G} \tilde\pi(x)\tilde\pi(r)
1_{e\in\tilde\gamma_{xr}}|\tilde\gamma_{xr}| & \text{from \eq{pi-ratio}}\\
& \leq  16a^2 \tilde\rho \tilde Q(e) & \text{from \eq{canonical-again}}\\
& \leq  16\theta R a^2 \tilde\rho Q(e) & \text{from
  \eq{Q-comparison}}
\ea

We conclude that our set of canonical flows achieves congestion
\be 
\rho \leq 16 \theta R a^2 \tilde\rho, \label{eq:good-congestion}\ee
albeit on a set of measure less than one, namely $\Omega_G$.
In the next section we will discuss the implications of this for mixing.

\subsection{Leaky random walks}\label{sec:leaky}
In this section we will prove \thmref{leaky-convergence-informal} by analyzing the general case of a substochastic random walk $P_G$ which is defined by the (unnormalized) restriction of the transition probabilities of a Markov chain $P$ to a subset $\Omega_G$ of its full state space $\Omega$.  Let $\Omega = \Omega_G \cup \Omega_B$ be a partition of $\Omega$, and define
\be
P_G(x,y) := P(x,y) 1_{x\in \Omega_G} 1_{y\in \Omega_G}.\label{eq:Prestricted}
\ee

If $P$ is ergodic then the stationary distribution $\pi$ has support on $\Omega_B$, so $\lim_{t\rightarrow\infty}P_G^t(x,\cdot) = 0$ for any starting point $x \in \Omega$.  However
there may be an intermediate range of times, $t_{\textrm{mix}} \leq t \leq t_{\textrm{fail}}$, for which $\|P_G^t(x,\cdot) - \pi\|_1 < \delta$ for some desired $\delta$, if the initial point $x \in \Omega_G$ is a \textit{warm start} taken from $x \sim \mu$ with $\|\pi - \mu\|_1$ sufficiently small. Define an $M$-warm distribution to be one satisfying $\mu(x)\leq M\pi(x)$ for all $x$.  Let $\Pi_G$ to be the projector onto $\bbR^{\Omega_G}$ , so that $P_G$ can be expressed as
\be P_G = \Pi_G P \Pi_G. \ee
To bound the gap of $P_G$ observe that the path comparison method (Thm 13.23 of \cite{peres-2008}) can be applied even to
substochastic matrices. Here it  implies that 
\be \Pi_G \cE\Pi_G \succeq_\pi \rho^{-1}\Pi_G. \label{eq:Pi-G-ineq}\ee
with $\rho$ from \eq{good-congestion}, and with $\succeq_\pi$ denoting
the semidefinite ordering with respect to the $\langle,\rangle_\pi$
inner product (i.e. $A\succeq_\pi 0$ means $\langle f, Af\rangle_\pi \geq
0$).  Eq.~\eq{Pi-G-ineq} does not directly give bounds on the gap of $P$.  Indeed we could
in principle have $P(x,x)=1$ for all $x\in\Omega_B$ in which case $P$
would have many eigenvalues equal to 1.  Thus without additional information we
cannot say anything more about $P$.

Instead we will analyze a slightly different Markov chain.  We will
define a ``filled-in'' chain $P_F$ by adding nonnegative numbers to
the entries of $P_G$ in such a way that $P_F$ is stochastic and has
$\pi$ as a stationary distribution.  This does not uniquely specify
$P_F$ and indeed $P$ also satisfies these conditions.  We will instead
choose $P_F$ in a way that guarantees fast mixing.  Specifically with
probability $1-\sum_{y\in\Omega} P_G(x,y)$ we will
forget our current location $x$ and jump according to some ``fill-in''
measure $\varphi$.  For this to result in $\pi$ being the stationary
distribution we must have $\varphi = \pi (I-P_G)$.  Defining the column vector ${\bf 1} := (1,1,\ldots,1)^T$, we have
\ba P_F &= P_G + {\bf 1} \varphi \\
\pi P_F  &= \pi(P_G + {\bf 1}\varphi )
 = \pi(P_G + {\bf 1}\pi(I-P_G) ) = \pi.
\ea

Now that $P_F$ is a proper Markov chain we will bound its Dirichlet
form.  Recall that in \eq{gap-def} we can assume that $f\perp_\pi
\mathbf{1}$ meaning that $\sum_x \pi(x)f(x)=0$.  Note as well that
\be\cE_F = I - P_F = I-P_G - \mathbf{1}\pi (I-P_G) = (I - \mathbf{1}\pi) (I-P_G).\ee
Define $D_\pi$ to have $\pi$ along the diagonal and zeros elsewhere.
Then 
\begsub{F-gap}
\langle  \cE_F f,f\rangle 
& = \langle (I-\mathbf{1}\pi) (I-P_G) f,f\rangle \\
& = f^T  (I-P^T_G) (I - \pi^T\mathbf{1}^T) D_\pi f \\
& = f^T  (I-P^T_G) D_\pi f & \text{since }f\perp_\pi\mathbf{1}\\
& = f^T (I-\Pi)D_\pi f + f^T  (\Pi -P^T_G) D_\pi f \\
& \geq f^T (I-\Pi)D_\pi f + \rho^{-1} f^T  \Pi D_\pi f &\text{using \eq{Pi-G-ineq}}  \\
& \geq \rho^{-1} f^T D_\pi f &\text{since }\rho\geq 1 \\
& = \rho^{-1}\Var_\pi[f] & \text{since }f\perp_\pi\mathbf{1}
\endsub

We conclude that $P_F$ mixes rapidly, while differing from $P_G$ only in
the events where leakage occurs.  More precisely we can define a
coupling between two processes: the first a walk evolving according to
$P_F$ and the second a walk in which $x$ moves to $y$ with probability
$P_G(x,y)$ and to a special state * with probability
$1-\sum_y P_G(x,y)$.  The state * is absorbing, meaning that when the
second walker is at state * it stays 
there.
Conditioned on the second walker not being in
state * the two walkers will be at the same location.  Additionally
the probability of the second walker ending in * equals the
probability that the first walker ever passed through $\Omega_B$.
If we start with a probability distribution $\mu$ and take $t$ steps then the probability
of ending in * is $\mu(P_F^t - P_G^t){\bf 1}$.   This quantity 
also equals the variational distance between the two walkers' probability distributions.

Note the warm start property is strictly
preserved by $P_F$ and $P_G$ since
\be \mu P_G^t \leq \mu P_F^t \leq M \pi P_F^t = M\pi.\label{eq:warm-preserved}\ee

In this case 
the probability that a path $x_1, \ldots,x_t$ (with $x_1\sim \mu$ and
$x_i\sim P_F(x_{i-1})$ for $i>1$) passes through $\Omega_B$ is
\be \leq \sum_{i=1}^t \Pr[x_i\in \Omega_B] \leq Mt\pi(\Omega_B),\ee
where we have used first the union bound and then \eq{warm-preserved}.
By the above arguments we have
\be\|\mu(P_F^t -P_G^t) \|_1 \leq Mt\pi(\Omega_B).\ee
Using the spectral gap of $P_F$ (from \eq{F-gap}) and the mixing bound
in \eq{mixing-gap} we conclude that
\be
\| \pi - \mu P_G^t \|_1  \leq Mt\pi(\Omega_B) + \pi_{\min}^{-1} e^{-  t/\rho} \label{eq:leaky-convergence}
\ee

We see that the leakage probability increases linearly with $t$ while
the usual distance to stationarity decreases exponentially with
$t$. In many cases this leaves a wide range of $t$ in which the RHS of
\eq{leaky-convergence} can be small.

\section{Efficient convergence of SQA for the spike cost function}\label{sec:spike-time}
In this section we apply the method from \secref{almost-canonical} with the easy-to-analyze chain $(\tilde{\pi},\tilde{P})$ taken to be the SQA Markov chain with heat-bath worldline updates for the system without the spike, and $(\pi,P)$ equal to the corresponding chain for the spike system.  

The subset $\Omega_{B}$ will be shown to satisfy $\tilde{\pi}(\Omega_B) \leq \cO(n^{-c})$ for a constant $c$ that we will choose so that we can prove a walker beginning in $\Omega_G$ is likely to mix before it hits a point in $\Omega_B$.   

\paragraph{Congestion of the spikeless chain.}  Recall that $\Omega_B$ is defined in terms of a set of canonical paths $\{\gamma\}$ on $\Omega$ with congestion $\tilde{\rho}$ for the spikeless chain, together with a subset $\Omega_{\theta}$ of points which are excluded from paths in $\{\gamma\}$ to obtain a new set of paths with congestion $\rho \leq \cO(\theta \tilde{\rho})$ for the chain with the spike, within the subset $\Omega_G$.   The spikeless distribution $\tilde{\pi}$ corresponds to a collection of $n$ non-interacting 1D ferromagnetic Ising models of length $L$ in the presence of 1-local fields that bias the distribution towards configurations of lower Hamming weight.  The spin-spin coupling is such that each broken bond in $x$ lowers $\tilde{\pi}(x)$ by a factor of $\Theta(\tanh(\omega))$.  

First we will bound the congestion of heat-bath worldline updates (defined in \secref{sqaPrelim}).  Here it is convenient to represent states $x\in \Omega$ by their worldlines $x = (\bar x_1 , \ldots , \bar x_n)$, where $\bar x_i := (x_{i,1},\ldots,x_{i,L})$. For the spikeless system $(\tilde \pi, \tilde P)$ spins in different worldlines do not interact and so the conditional distribution of the $i$-th worldline is equal to the marginal of the stationary distribution on that worldline,
\be
\tilde\pi(\bar{x}'_i |\bar{x}'_1,\ldots,\bar{x}'_{i-1},\bar{x}'_{i+1},\ldots,\bar{x}'_n)= \sum_{\bar{x}_j : j \neq i} \tilde\pi(\bar{x}_1,\ldots,\bar{x}'_i,\ldots,\bar{x}_n)
\ee
Including a $1/2$ probability of the chain not moving at each step in order to make it irreducible, the probability of a transition $P\left((\bar z_1,\ldots,\bar z_i, \ldots , \bar z_n),(\bar z'_1,\ldots,\bar z'_i, \ldots , \bar z'_n)\right)$ that updates the $i$-th worldline ($\bar z_j = \bar z_j' $ for all $j \neq i$) is
\be
P\left((\bar z_1,\ldots,\bar z_i, \ldots , \bar z_n),(\bar z'_1,\ldots,\bar z'_i, \ldots , \bar z'_n)\right) = \frac{1}{2n} \sum_{\bar{z}_j : j \neq i}\tilde\pi(\bar{z}_1'',\ldots,\bar{z}'_i,\ldots,\bar{z}''_n)
\ee

The path $\gamma_{xy}$ from $x = (\bar{x}_1,\ldots,\bar{x}_n)$ to $y = (\bar{y}_1,\ldots,\bar{y}_n)$ proceeds by updating the worldlines in order $\{1,\ldots,n\}$.  The paths have length $|\gamma_{xy}| = n$.  The $k$-th step of the path $\gamma_{xy}$ will go through the edge $(z,z')$ with $z = (\bar{y}_1,\ldots,\bar{y}_{k-1},\bar{x}_k,\ldots,\bar{x}_n)$ and $z' = (\bar{y}_1,\ldots,\bar{y}_k,\bar{x}_{k+1},\ldots,\bar{x}_n)$.  To evaluate the sum \eq{compare-complete} we apply the standard encoding trick~\cite{sinclair-1992}.  Define an injective function $\eta_{(z,z')}$ which maps the paths through $(z,z')$ into the state space $\Omega$,
$$
\eta_{(z,z')}(\gamma_{xy}) = (\bar{x}_1,\ldots,\bar{x}_{k-1},\bar{y}_{k},\ldots,\bar{y}_n),
$$
which is injective since $z$ and $\eta_{(z,z')}(\gamma_{xy})$ provide sufficient data to uniquely determine $\gamma_{xy}$.  Notice that $\tilde\pi(x)\tilde\pi(y) = \tilde\pi(z)\tilde\pi\left(\eta_{(z,z')}(\gamma_{xy})\right)$, and so the congestion $\tilde \rho(z,z')$ is

\ba
\tilde \rho(z,z') &= \frac{1}{\tilde\pi(z)P(z,z')} \sum_{\gamma_{xy} \ni (z,z')} \tilde\pi(x)\tilde\pi(y) |\gamma_{xy}|\label{eq:enc1} \\
& = \frac{n}{\tilde P(z,z')} \sum_{y_{xy}\ni (z,z')} \tilde\pi \left(\eta_{(z,z')}(\gamma_{xy}) \right)\label{eq:enc2}\\
&= \frac{n}{\tilde P(z,z')} \sum_{\substack{\bar{x}_1,\ldots,\bar{x}_{k-1} \\ \bar{y}_{k+1},\ldots,\bar{y}_n}}\tilde\pi(\bar{x}_1,\ldots,\bar{x}_{k-1},\bar{y}_{k},\ldots,\bar{y}_n)\label{eq:enc3}\\
&= 2 n^2 \label{eq:enc4},
\ea
where in going from \eq{enc2} to \eq{enc3} we do not sum over $\bar{y}_k$ because it is fixed by $z'$.  Finally, since the edge $(z,z')$ was arbitrary we have 
\be
\tilde \rho = \cO(n^2).\label{eq:tildeRho}
\ee
Now we will analyze the single-site Metropolis chain for the spikeless system $(\tilde \pi, \tilde P_M)$.  The path from $x = ((x_{1,1},\ldots,x_{1,L}),\ldots,(x_{n,1},\ldots,x_{n,L}))$ to $y = ((y_{1,1},\ldots,y_{1,L}),\ldots,(y_{n,1},\ldots,y_{n,L}))$ proceeds as follows: for each $i$ in order from $\{1,\ldots,n\}$ update spin $(i,j)$ for $j$ going in order from $\{1,\ldots,L\}$.  The paths have length $|\gamma_{xy}| = n L$.   Any edge $(z,z')$ along such a path will create at most two new broken bonds (i.e. pairs of spins which disagree) along the direction $\{1,\ldots,L\}$, and so
\be
\tilde P_M(z,z')= \frac{1}{2 n L} \min\left\{1,\frac{\tilde \pi(z')}{\tilde \pi(z)}\right\} = \Omega\left((n L)^{-1}\tanh(\omega)\right) \label{eq:pmtb}
\ee

Just as for the heat-bath worldline updates, we apply an encoding function to injectively map the paths through an arbitrary edge into the state space.  The only difference from the version above is that $z$ and $\eta_{(z,z')}\left(\gamma_{xy}\right)$ may in the worst-case each have two broken imaginary-time bonds which are not present in either $x$ or $y$, and so 
\be
\tilde \pi(x) \tilde \pi(y) = \cO\left(\coth(\omega)^2 \tilde \pi(z) \tilde \pi\left(\eta_{(z,z')}(\gamma_{xy})\right)\right)\label{eq:mea}.
\ee
Applying the same calculations in \eq{enc1}-\eq{enc3} but with \eq{pmtb} and \eq{mea}, and using the fact that $\coth(\omega) = \cO(\omega_{\min}^{-1}) = \cO(n L \beta^{-1})$, the congestion is
\be
\tilde \rho_M = \cO(L^5 n^5 \beta^{-3}) =  \cO(n^{15}\beta^{-9/2}). \label{eq:tildeRhoM}
\ee
Comparing \eq{tildeRhoM} with \eq{tildeRho} shows that there is a large polynomial overhead resulting from single-site updates, which arises from the strong interactions that occur in the imaginary-time direction when the transverse field term of the QA Hamiltonian is small.  Nevertheless, the bound \eq{tildeRhoM} is still efficient in the sense of polynomial time, and will suffice to obtain an overall polynomial run time for single-site Metropolis chain of the spike system.

\paragraph{$\Omega_{\theta}$ and the spike time distribution.}The states in $\Omega_{\theta}$ which will be excluded from the set of paths $\{\gamma\}$ are those which have $|x_i| \in I_S:= (n/4 - n^\zeta/2, n/4 + n^\zeta/2)$ for too many $i$.  Define $1_S:\{0,1\}^n \rightarrow \{0,1\}$ to be the indicator function for the spike i.e. $1_S(z) = 1$ if $z \in I_S$, and $1_S(z)$ is zero otherwise.  The spike time for $x \in \Omega$ is defined to be
\be
\ST(x) = \sum_{i=1}^L 1_S(x_i).
\ee
Let $\epsilon := \frac{1}{2} - \alpha$, and define
\be
\Omega_{\theta} = \left\{x \in \Omega : \ST(x) \geq \frac{L}{n^{\frac{1}{2}(1-\epsilon) - \zeta}}\right\}.
\ee
Set $\beta  = n^{\epsilon/2}$ so that every $x \notin \Omega_{\theta}$ satisfies
\be
\frac{\pi(x)}{\tilde \pi(x)} \geq \left(\frac{\tilde Z}{{Z}} \right) \exp\left[-\frac{\beta n^{\alpha}}{L}\cdot\left(\min_{x\notin \Omega_{\theta}} \ST(x)\right) \right ] = \cO(1), \label{eq:thetaSmall}
\ee
which shows $\theta = \cO(1)$ in the congestion bound \eq{good-congestion}.  The remainder of the section will be devoted to computing the $m$-th moment of the random variable $\ST \sim \tilde{\pi}$, with $m = c/\epsilon$, in order to show,
\be
\Pr\left[\ST \geq \frac{L}{n^{\frac{1}{2}(1-\epsilon) - \zeta}} \right]_{\tilde{\pi}} \leq \cO(n^{-c}), \label{eq:omegaThetaSmall}
\ee
which is equivalent to the statement $\tilde{\pi}(\Omega_{\theta}) \leq \cO(n^{-c})$.  

To calculate the moments $\langle \ST^m \rangle_{\tilde{\pi}}$ we will relate them to expectation values of the spikeless quantum system, and use the fact that the latter is exactly solvable because the qubits are non-interacting.    Recall that the quantum expectation value of an operator $A$ can be expressed as a derivative of the partition function,
\begin{equation}
\langle A \rangle =\frac{1}{\mathcal{Z}} \tr\left[A e^{-\beta H}\right] = \frac{1}{\mathcal{Z}}\frac{\partial}{\partial \lambda} \tr \left[e^{-\beta H + \lambda A}\right] |_{\lambda = 0}= \frac{1}{\mathcal{Z}(0)}\frac{\partial \mathcal{Z}(\lambda)}{\partial \lambda}|_{\lambda = 0}.
\end{equation}
Passing from $\mathcal{Z}(\lambda)$ to the corresponding Suzuki-Trotter approximation $Z(\lambda)$ in the above expression changes the value by at most a multiplicative factor of $(1 \pm \mathcal{O}(L^{-1})$ (a proof of this fact will appear in a future work~\cite{Crosson-2016}). 
Let $\{|k\rangle : k = 0,\ldots,n\}$ be a basis of states for the symmetric subspace which are labeled by Hamming weight, and let $S = \sum_{k\in I_S}|k\rangle \langle k|$.  Since the observable $S$ is diagonal in the computational basis we can include the term $\lambda S$ into the diagonal part of the Hamiltonian for the quantum-to-classical mapping and compute,  
\begin{align}
\langle S \rangle_{\tilde{\sigma}} &= \frac{1}{Z}\frac{\partial}{\partial \lambda}\left [ \sum_{x_1,\ldots,x_L}  \exp\left(\sum_{i=1}^L -\frac{\beta \tilde{f}(x_i)}{L} + \frac{\lambda \langle x_i | S |x_i\rangle}{L}\right)\prod_{j=1}^{n} \phi(\bar{x}_j) \right]|_{\lambda = 0}\\
&= \frac{1}{Z}\sum_{x_1,\ldots,x_L} \left[\frac{1}{L}\sum_{i=1}^L 1_S(x_i) \right] e^{ -\frac{\beta}{L}\sum_{i=1}^L \tilde{f}(x_i)}\prod_{j=1}^{n} \phi(\bar{x}_j)\\[5pt]
&= \sum_{x\in \Omega} L^{-1}\ST(x) \tilde{\pi}(x) = L^{-1}\langle \ST \rangle_{\tilde{\pi}}
\label{eq:expectS}\end{align}
Since the low-temperature thermal state and the ground state have a large overlap, the expectation value \eq{expectS} can be computed within the ground state.  To compute the error caused by this replacement we will examine the density of states of the spikeless system.   First add a constant shift to the Hamiltonian so that $\tilde H|\tilde{\psi}_0\rangle = 0$.   Next, let $|\tilde\psi_1\rangle,\ldots,|\tilde \psi_n\rangle$ denote the excited eigenstates of $\tilde H$. Define $\Delta := 2\sqrt{(1-s)^2 + s^2}$ and observe that $|\tilde{\psi}_k\rangle$ is an eigenstate of $\tilde H$ with eigenvalue $k\Delta$, and that the degeneracy of the $k$-th energy level is $\binom{n}{k}$ so  
\ba
\|\tilde{\sigma} - |\tilde{\psi}_0\rangle \langle \tilde{\psi}_0|\|_1 \leq \sum_{k = 1}^{n} e^{-\beta \Delta k} \binom{n}{k} = \cO(n e^{-n^\epsilon}), \label{eq:thermalReplaceError}
\ea
and since $\epsilon > 0$ is a constant this error will be sub-leading.

Since the spikeless system is non-interacting the ground state can be written explicitly, and the ground state probability distribution on the Hamming weights is a binomial distribution~\cite{Reichardt-2004}.  It follows that $\langle S \rangle_{\tilde{\sigma}}$ is asymptotically never larger than the central binomial coefficient times the width $n^\zeta$ of the spike term, therefore
\be
\langle \ST \rangle_{\tilde{\pi}} = \cO\left(L n^{\zeta - 1/2} \right). \label{eq:centralBinomial}
\ee
To obtain \eq{omegaThetaSmall} we will use the moment inequality,
\be
\Pr\left[\ST  \geq b \right]_{\tilde{\pi}} \leq \frac{\langle \ST^m\rangle_{\tilde{\pi}}}{b^m},\label{eq:moment-bound}
\ee
with $b = L n^{-\frac{1}{2}(1-\epsilon)}$.   By definition we have,
\begin{align}
\langle \ST^m\rangle_{\tilde{\pi}} &= \sum_{z_1,\ldots,z_L}   \tilde{\pi}(z_1,\ldots,z_L) \left( \sum_{t = 1}^L 1_S(z_t) \right )^m \\
&= \sum_{z_1,\ldots,z_L}  \tilde{\pi}(z_1,\ldots,z_L)  \sum_{t_1,\ldots,t_m}^L 1_S(z_{t_1})\ldots1_S(z_{t_m})  \\
&= \sum_{t_1,\ldots,t_m}^L \langle 1_S(z_{t_1})\ldots1_S(z_{t_m}) \rangle_{\tilde{\pi}} \label{eq:sumtime}.
\end{align}
To compute these $m$-point correlation functions we return to the quantum description (this generates a multiplicative error of size $1 \pm \cO(L^{-1})$, which will make a sub-leading contribution to the $m$-th moment and thus will be ignored),  
\begin{equation}
\langle 1_S(z_{t_1})\ldots1_S(z_{t_m}) \rangle_{\tilde{\pi}} = \langle  e^{-\tau_1 H} S e^{-(\tau_2 - \tau_1)H} S \ldots e^{-(\tau_m - \tau_{m-1})H} S e^{-(\beta - \tau_m) H} \rangle_{\tilde{\sigma}}
\end{equation}
where $\tau_i := \beta t_i / L$.  Once again we replace the low-temperature thermal state with the ground state and incur a sub-leading error as in \eq{thermalReplaceError}.  Therefore,
\be
\langle\ST^m\rangle_{\tilde{\pi}} = \sum_{t_1,\ldots,t_m}^L \langle \tilde{\psi}_0|e^{-\tau_1 H} S e^{-(\tau_2 - \tau_1)H} S \ldots e^{-(\tau_m - \tau_{m-1})H} S e^{-(\beta - \tau_m) H} |\tilde{\psi}_0\rangle \label{eq:STm72}
\ee
Since the ground state, the Hamiltonian, and the operator $S$ are all bit-symmetric, the expectation can be evaluated in the symmetric subspace.  Expanding each of the terms in \eq{STm72} in the basis of symmetric energy eigenstates $\{|\tilde{\psi}_k\rangle\}$,
\begin{equation}
\langle\ST^m\rangle_{\tilde{\pi}} =  \sum_{\substack{k_1,\ldots,k_m \\ t_1,\ldots,t_m}}  e^{-(\tau_2 - \tau_1) \Delta{k_1}} \cdots e^{-(\tau_m - \tau_{m-1})\Delta{k_1}} \langle \tilde{\psi}_{0}| S|\tilde{\psi}_{k_1}\rangle \langle \tilde{\psi}_{k_2}|S|\tilde{\psi}_{k_3}\rangle \cdots \langle \tilde{\psi}_{k_m} | S| \tilde{\psi}_{0}\rangle \label{eq:energyExpand3}
\end{equation}
States with higher energy will contribute less to the sum over all times $t_1,\ldots, t_m$ in \eq{STm72}, because the exponentials in \eq{energyExpand3} decay more quickly.  
For $k_i > 0$, the sum over $t_i$ can be truncated whenever $\tau_i - \tau_{i-1} \gg 1/k_i\Delta$.

Since the ground state wave function is a binomial distribution the mean spike time will only be large when the peak of the ground state is near the support of the spike $I_S$.  In the range of the adiabatic parameter in which this occurs the excited spikeless eigenstates satisfy $\langle \tilde{\psi}_i|k\rangle \leq |\langle\tilde{\psi}_0|k\rangle|\leq \cO(n^{-1/4})$ for all $i = 1,\ldots,n$ and $k\in I_S$, because the ground state wave function is centered on the spike and the excited state wave functions have a greater spread, which can be seen from the explicit form of the spikeless eigenfunctions given in ~\cite{Kong-2015}.
Now we define $g_i := t_i - t_{i-1}$ and relabel the sum of $t_1, \ldots ,  t_m = 0,\ldots,L$ by a sum over the $g_i$.  For the purpose of obtaining an upper bound on the $m$-th moment we relax the constraint $\sum_i g_i = L$, and instead sum over the full range $g_i = 1,\ldots,L$ for each $i$.  
Using these facts we can upper bound \eq{energyExpand3} by
\ba
\langle\ST^m\rangle_{\tilde{\pi}} 
& \leq
n^{m(\zeta -1/2)}
\sum_{\substack{k_1,\ldots,k_m \\ g_1,\ldots,g_m}}  e^{-g_1 \Delta{k_1}} \cdots e^{-g_m\Delta{k_m}}
\label{eq:ST-ub}
\ea
We will now organize the terms of \eq{ST-ub} according to the number $\ell$ of excited energies  $\tilde{E}_{k_i} > 0$ they contain.  There are $\binom{m}{\ell}$ terms of \eq{ST-ub} that contain $\ell$ eigenstates above the ground state, and for each $\ell$ we must sum over the $g_{a_1},\ldots,g_{a_\ell}$ for which the corresponding $k_{a_1},\ldots,k_{a_\ell}$ are non-zero.  Now \eq{ST-ub} becomes 
\begin{align}
\langle\ST^m\rangle_{\tilde{\pi}}
& \leq n^{m(\zeta - 1/2)}
 \sum_{\ell = 1}^m \binom{m}{\ell}  \frac{L^{m-\ell}}{(m-\ell)!} 
\sum_{\substack{g_{a_1}, \ldots g_{a_\ell} \\ k_{a_1} \ldots k_{a_\ell}}} \prod_{i = 1}^\ell e^{-g_{a_i}\Delta{k_{a_i}}} 
\end{align}
where the factor of $L^{m-\ell}/(m-\ell)!$ results from performing the sum over the $m - \ell$ of the $g_i$ which have $k_i = 0$. Now we sum over $g_{a_1},\ldots,g_{a_\ell}$ using the fact that $\sum_{g = 1}^L   e^{-g k} \leq k^{-1}$,
\begin{equation}
\langle \ST^m \rangle_{\tilde{\pi}} \leq  L^m n^{m(\zeta - 1/2)} \sum_{\ell = 0}^m \binom{m}{\ell} (\beta \Delta)^{-\ell} \sum_{k_{a_1} \ldots k_{a_\ell}}^n \prod_{i = 1}^\ell  \frac{1}{k_i}   .
\label{eq:m-moment}\end{equation}
Using $\binom{m}{\ell} \leq m^{\ell}$ and $\sum_{k = 1}^n \leq \log(n) + 1$, at last this becomes
\begin{equation}
\langle \ST^m \rangle_{\tilde{\pi}} \leq  L^m n^{m(\zeta-1/2)} \sum_{\ell = 0}^m  \left(\frac{\log(n) + 1}{m \beta \Delta}\right)^{\ell} 
\end{equation}
Since $m$ and $\Delta$ are constant and $\beta = n^{\epsilon/2}$ with fixed $\epsilon > 0$ the terms with inverse powers of $\beta$ are sub-leading and so $\langle \ST^m \rangle_{\tilde{\pi}} \leq \cO(L^m n^{m(\zeta - 1/2)})$.  Finally, applying \eq{moment-bound} with $m = c/\epsilon$ yields the desired result \eq{omegaThetaSmall}.

\paragraph{Adiabatic schedule.} Here we show that a discretization of the adiabatic path with $1/\poly(n)$ step size is sufficient to fulfill the
statement we need for the warm starts in \secref{leaky}. We will take the largest value of the adiabatic parameter to be $s_{\max} = 1 - n^{-1}$ so that $||\psi_0(1)\rangle - |\psi_0(s_{\max})\rangle|_1 \leq \poly(n^{-1})$, and the global minimum of the cost function can be obtained by sampling from $\Pi$ at $s = s_{\max}$ with essentially the same probability at it would be obtained by sampling from the ground state probability distribution. 

We will sample from $\pi$ at several values of the adiabatic parameter $s_1,\ldots,s_{\max}$.  Define $s_i  := s_0 - i \Delta s$, $\omega_i := \beta (1-s_i)/L$, $\Delta \omega := \beta \Delta s / L$, and let $\pi^i$ be the stationary distribution \eqref{eq:pimain} when the adiabatic parameter is $s_i$.   At each stage we simulate the Markov chain \eq{Metropolis} for sufficiently many steps to achieve a variational distance to the stationary distribution of $\exp(-n^{\Omega(1)})$. These errors then add up to a negligible amount.  To choose a step size $\Delta s$ satisfying the warm start condition $\pi_{i+1} \leq 2 \pi_i$ we'll use a claim which is inspired by Lemma 5.1 of \cite{bravyi-2008} but is a bit simpler in the classical case.

\begin{lemma}
\label{lem:stepsize}  Let $E_1, E_2:A\rightarrow\mathbb{R}$ be energy functions on a domain $A$ and define $Z_i := \sum_{x\in A} e^{-E_i(x)}$ and $p_i(x) = e^{-E_i(x)}/Z_i$ for $i=1,2$.  If $\max_x|E_1(x)-E_2(x)|\leq \delta$, then $|\log(Z_1/Z_2)|\leq \delta$ and
$\max_x |\log(p_1(x)/p_2(x))|\leq 2\delta$. 
\end{lemma}
\par\noindent\textit{Proof.}
Applying the uniform bound $| E_2(x) - E_1(x) | \leq \delta$ for all $x\in A$ to the sum $Z_2 = \sum_{x\in A} e^{-E_1(x)}$ leads to $e^{-\delta}Z_1 \leq Z_2 \leq e^{\delta} Z_1$, therefore $|\log(Z_1/Z_2)|\leq \delta$.  Since $|\log(p_1(x)/p_2(x))| = |\log(p_2(x)/p_1(x))|$, we treat the case $p_1(x) \geq p_2(x)$ (which implies $E_2(x) \geq E_1(x)$), while the other case follows similarly.  
\ba
\max_x |\log(p_1(x)/p_2(x))| &= \max_x |\log(Z_2/Z_1) + \log(e^{-(E_1(x) - E_2(x))})| \nonumber \\
 &= |\log(Z_2/Z_1)| + \max_x|\log(e^{E_2(x)-E_1(x)})| \nonumber\\
&\leq 2 \delta \nonumber
\ea

Applying \lemref{stepsize} with the form of the stationary distribution \eq{pimain} we may take
\[
\delta  = \frac{\beta}{L} \Delta s f_{\max} + n L \log \left(\frac{ \tanh(\omega_i - \Delta\omega)}{\tanh(\omega_i)}\right) = \cO\left(\beta \Delta s \; n \log(n)\right),
\]
therefore taking $\Delta s = \cO\left((\beta n \log(n))^{-1}\right)$ fulfills the warm start condition.

\paragraph{Quasi-stationary mixing.} In this section we will bound the overall run time of SQA applied to the spike cost function.  First we need to show that taking $c$ in \eq{omegaThetaSmall} to be a sufficiently large constant will allow the leaky walk for the spike system sample from $\Omega_G$ according to the quasi-stationary distribution $\pi$ for an expected time of $n^q$, for any desired constant $q$, before it is eventually likely to escape into $\Omega_B$.  

Inserting $\tilde \rho$ from \eq{tildeRho} into \eq{good-congestion} yields $\rho = \cO(n^2)$.   To apply \eq{leaky-convergence} we next must consider $\pi^{-1}_{\min}$.  From \eq{pimain} we have $\log \pi_{\min}^{-1} \leq \cO(n L \log(n))$ because there can be $L$ pairs of bits which disagree in each of the $n$ worldlines.  However, according to \eq{pimain} these disagreements follow a binomial distribution with mean $\beta (1-s)$, and so we may abort the algorithm with exponentially small probability if it ever encounters a configuration with $\Omega(\beta \log n)$ jumps in any worldline, which allows us to take $\log\pi^{-1}_{\min} = \cO(n \beta \log(n)$.     This implies a mixing time of $t_{\textrm{mix}} = \cO(n^3 \beta \log n)$ within $\Omega_G$ for the SQA spike chain with single qubit worldline updates at each value $s_i$ of the adiabatic path.

Meanwhile, from \eq{ThetaToBee} together with \eq{omegaThetaSmall} we have $\pi(\Omega_B) \leq \Theta(\tilde{\rho} \tilde{\pi}(\Omega_{\theta}))= \Theta(n^{2 - c})$.   At each step $s_i$ of the adiabatic path, after time $t \geq t_{\textrm{mix}}$ the leaky random walk mixes to within a distance $\cO(t \pi(\Omega_B))$ so by \eq{leaky-convergence} it suffices to take $c = 2 + q + \log(1/\delta)$ in order for the leaky walk to be with distance $\delta$ to the stationary distribution $\pi$ for times $t_{\textrm{mix}} \leq t \leq \Theta(n^q)$.  

Finally, since there are $\tilde \cO(n \beta)$ steps of the adiabatic path, and each worldline update takes $\tilde \cO(n^2 \beta^2)$ time to implement, we obtain a total run time of $\tilde \cO(n^{6}\beta^4)$, which implies the $\tilde \cO(n^7)$ stated in Theorem 1.   Repeating the analysis above for single-site Metropolis updates, we combine the congestion \eq{tildeRhoM} with $\log \pi^{-1}_{\min} = \tilde \cO(n \beta)$ and $\tilde \cO(n \beta)$ steps of the adiabatic path to arrive bound the total run time by $\tilde \cO(n^{17})$ as stated in Theorem 1.  

\section*{Acknowledgments}
We thank Dave Bacon, Wim van Dam and Alistair Sinclair for helpful conversations.   Elizabeth Crosson gratefully acknowledges funding provided by the Institute for Quantum Information and Matter, an NSF Physics Frontiers Center (NSF Grant PHY-1125565) with support of the Gordon and Betty Moore Foundation (GBMF-12500028), and is also grateful for support received while completing a portion of this work at the MIT Center for Theoretical Physics with funding from NSF grant number CCF-1111382.
Aram Harrow was funded by NSF grants CCF-1111382 and CCF-1452616 and ARO contract
W911NF-12-1-0486.

\bibliographystyle{hyperabbrv}

\end{document}